\documentclass[12pt,showpacs,preprintnumbers,superscriptaddress,amsmath,amssymb,nofootinbib]{revtex4-1}
\usepackage{graphicx,hyperref}
\usepackage{dcolumn}
\usepackage{bm}
\usepackage{amssymb}
\usepackage{amsmath}
\usepackage{epsfig}    
\usepackage{color}
\usepackage{slashed}
\usepackage{hhline}

\def\be{\begin{equation}}
\def\ee{\end{equation}}
\newcommand{\bea}{\begin{eqnarray}}
\newcommand{\eea}{\end{eqnarray}}

\numberwithin{equation}{section}

\begin{document}

\title{ Constraining Effective Self Interactions of Fermionic Dark Matter}

\author{Kamakshya Prasad Modak}
\email{kamakshya.modak@saha.ac.in}
\affiliation{Astroparticle Physics and Cosmology Division, Saha Institute of Nuclear Physics, Kolkata 700064, India}


\begin{abstract}

The idea of Dark Matter (DM) with self interaction was invoked to resolve a number of discrepancies between the simulation based predictions by collisionless cold DM and the astrophysical observations on galactic and subgalactic scales. Evidences for self interaction would have striking implications for particle nature of DM. In order to reconcile such astrophysical observations for self interaction with particle properties for DM, we consider the general scenario of self interacting Dirac fermionic DM, $\chi$. Also since the exact particle physics model for DM is yet to be probed, we simply adopt the effective model independent framework for DM self interaction which occurs via the most general effective 4-fermion operators invariant under both Lorentz and CPT transformations. From the thorough investigation of the interrelations among the parameters in this framework, namely, the effective DM self couplings ($G_{i}$), DM mass ($m_{\chi}$) and relative velocity ($v_{\rm rel}$), it can be inferred that $G_{i}
$ decrease with increasing $m_{\chi}$ for a given DM self interaction strength. Moreover, for few types of effective operators the values of $G_{i}$ fall off with increasing $v_{\rm rel}$ while they remain roughly constant for a wide range of $v_{\rm rel}$ for other cases. In addition, the parameter space in this framework is constrained by the claimed observational results of ${\sigma \over m_{\chi}}$ on cluster scales (Abell 3827, Bullet Cluster) after averaging the DM self interaction cross sections over DM velocity distribution in the cluster. This puts interesting constraints on the values of effective DM self couplings for different fermionic DM masses for various effective operators (scalar, vector, etc.) of DM self interactions in this scenario. Some other implications of DM effective self interaction are also discussed in this model independent framework.

\end{abstract}
\maketitle
\newpage

\section{Introduction \label{intro}}

It is now very well established from various astrophysical and 
cosmological evidences~\cite{Begeman:1991iy, Komatsu:2010fb, Massey:2007wb}
that a large fraction of the universe's mass consists of
unknown non-luminous matter, namely, Dark Matter (DM). The Planck 
satellite-borne experiment~\cite{Ade:2013lta} estimates
from the anisotropy study of the expected isotropic nature of 
Cosmic Microwave Background Radiation (CMBR) that 
$\sim$ 26.8\% of the total energy content of the universe is dominated by DM. Since there is
no candidate of DM in the Standard Model (SM) of particle physics, it is believed that there exists
a new physics sector beyond the SM. Although the nature of DM remains unknown still now, it is
conjectured that the weakly interacting massive particles (WIMPs) being stable, non-relativistic,
electrically neutral and colourless are very well-motivated candidates of DM. 
The mass of WIMPs can vary from a few GeV to TeV with suitable strength of interaction to yield 
observed DM relic abundance depending on various models of particle physics~\cite{scalar_singlet, 
veltman, Burgess:2000yq, kaluza, triplet, Modak:2012wk, smssm, axion, 
sing_ferm, idm, Cirelli, Hooper:2012sr, Modak:2013jya, Modak:2014vva, Modak:2015uda}. 

On the other hand, the formation and evolution of structures in the universe 
can also predict precise nature of DM sector. 
The large scale structure of the universe can almost be successfully accounted for in case of
collisionless cold dark matter (CCDM) paradigm. But such scenario fails in explaining the small-scale
structure of the universe. There are persistent problems such as {\it core-vs-cusp} problem~\cite{deBlok:2009sp}, 
{\it missing satellite} problem~\cite{Bullock:2010uy}, {\it too-big-to-fail} problem~\cite{BoylanKolchin:2011de} etc. which 
arise in the CCDM scenario while 
predicting the precise nature of the small scale structures in the universe.
The central density profiles of the DM halos surrounding the dwarf galaxies indicate flat
cores from the observation of galactic rotation curves~\cite{Flores:1994gz,de Blok:2001fe,
Simon:2004sr,Oh:2010ea,deNaray:2011hy} while numerical simulations considering
CCDM model predict the steep cusps~\cite{Dubinski:1991bm,Navarro:1996gj,Wechsler:2001cs}. 
Similar discrepancy has also been noted for the case of
observations of galaxy clusters~\cite{Sand:2003ng,Newman:2009qm,Newman:2012nw}. 
This discrepancy known as the {\it core-vs-cusp} problem cannot be
resolved by effects such as supernova feedback etc. Another problem, namely, the {\it missing
satellite} problem is the discrepancy of $\sim \cal{O}$(10) magnitude 
between the expected and the observed 
numbers of satellites within the Milky Way~\cite{Kauffmann:1993gv,Klypin:1999uc,
Moore:1999nt,Bullock:2010uy}. Although recently the Sloan Digital Sky Survey
has discovered many faint galaxies with much precision, 
there has been a multiplying factor of $\sim 5-20$
discrepancy between the numbers of known and undiscovered dwarf galaxies~\cite{Tollerud:2008ze,
Walsh:2008qn,Bullock:2009gv}. Therefore there exists 
an unresolved issue on the predicted number of Milky Way subhalos in CCDM scenario. 
Moreover, there exists another interesting discrepancy whereby the estimations of 
numerical values of host subhalo mass and density become so big that they fail to confront 
the observational evidences, known as the {\it too-big-to-fail} problem. 
The largest velocity dispersions of the brightest Milky Way dwarf spheroidal (dSph) galaxies observed
from the rotation curves of these galaxies may indicate that these galaxies are hosted
by the largest subhalos in the Milky Way halo. But the central densities of such highly massive host
subhalos as predicted by the CCDM-only simulations are too large and also the masses of such subhalos
become too huge to host those above-mentioned brightest dwarf satellite 
galaxies~\cite{Strigari:2007ma,BoylanKolchin:2011de,BoylanKolchin:2011dk}.
The maximum circular velocities of these satellite galaxies (or dSphs) are observed to be much less
lower than those estimated by numerical simulations in CCDM paradigm. This {\it too-big-to-fail} problem
may be attributed to have similar resolution like that of the {\it core-vs-cusp} problem since 
such massive subhalos can be made consistent with the observed dSphs if the central densities are
reduced. 

With these long-established discrepancies involved in the CCDM paradigm, a suitable modification
of such scenario is much required. Inclusion of warm DM~\cite{Bode:2000gq}, i.e., DM with some kinetic energy was the 
first attempt towards such modification to resolve the missing satellite problem. However, the
other severe problems cannot be fully resolved with this warm DM~\cite{VillaescusaNavarro:2010qy,
Maccio:2012qf}. There are several valid attempts~\cite{Carlson:1992fn, Spergel:1999mh, Kaplinghat:2000vt, 
Boehm:2000gq, Sigurdson:2003vy, Hooper:2007tu, Kaplan:2009de, Chu:2014lja} 
to alleviate such problems. These include the
proposition of self-interacting DM as a solution to all such discrepancies~\cite{Carlson:1992fn, Spergel:1999mh}. 
The energy transfer from
the outer hotter region in the halo to the central colder region can form a core structure 
within a central region of the halo in self-interacting DM scenario. 
Also, in this case the number of Milky Way satellite galaxies is much reduced due to the collisional
stripping of the subhalos~\cite{Wandelt:2000ad}. 

Since the beginning of the proposition of the self-interacting DM, 
various numerical simulations~\cite{Dave:2000ar, Vogelsberger:2012ku, Rocha:2012jg,
Peter:2012jh, Zavala:2012us} have tried
to estimate its value for different astrophysical observations. 
Early simulation had predicted ${\sigma \over m} \sim 1-10$ cm$^2$/g 
required to yield the observed flattening of the
central densities of dwarf galaxies~\cite{Dave:2000ar}. 
Moreover, the abundance of Milky Way subhalos is 
considerably reduced for ${\sigma \over m} \sim 10$ cm$^2$/g.
Other simulation~\cite{Yoshida:2000uw} had found that the value of ${\sigma \over m}$ to be $\geqslant 0.1$ cm$^2$/g
to avoid core formation in halos of galaxy clusters, which violates the
gravitational lensing observational constraints for the cluster CL 0024+1654.
The parameter space of ${\sigma \over m} \sim 0.3 - 10^4$ cm$^2$/g is ruled out requiring the non-evaporation 
constraints of elliptical galaxy halos within hot cluster halos~\cite{Gnedin:2000ea} whereas the cluster ellipticity
limits require ${\sigma \over m} \geqslant 0.02$ cm$^2$/g~\cite{MiraldaEscude:2000qt}. 
More recent numerical study on clusters, however, have found no 
significant evaporation of subhalos for ${\sigma \over m} \sim 1$ cm$^2$/g~\cite{Rocha:2012jg,Peter:2012jh}.
The current simulations~\cite{Vogelsberger:2012ku,Rocha:2012jg,Peter:2012jh} hint
the upper and lower limits of self-interaction as ${\sigma \over m} \sim 0.1 - 10$ cm$^2$/g to 
resolve the core-vs-cusp and too-big-to-fail problems on dwarf scales. 
Although there are controversies~\cite{Zavala:2012us} regarding such
predicted lower limit of self-interaction (${\sigma \over m} \sim 0.1$ cm$^2$/g).
On the other hand, the observational constraints from cluster and Milky Way scales
restrict the self-interaction as ${\sigma \over m} \geqslant 0.1 - 1$ cm$^2$/g~\cite{Vogelsberger:2012ku,
Rocha:2012jg,Peter:2012jh}.

In a very recent analysis~\cite{Massey:2015dkw}, it is claimed from the observation of a particular galaxy in the cluster
Abell 3827 with the Hubble space telescope imaging that an interesting offset of $\sim 1.62^{+0.47}_{-0.49} $ kpc (68\% CL)
between its DM and its stars (normal baryonic matter) is observed. This offset is thought to be
a consequence of self-interaction of DM (${\sigma \over m} \sim (1.7\pm0.7)\times 10^{-4}$ cm$^2$/g), 
and a lower bound on $\sigma \over m$, ${\sigma \over m} \geqslant 10^{-4}$ cm$^2$/g
is given from their analysis~\cite{Massey:2015dkw}. 
On the other hand, using a very 
different kinematic analysis for the same galaxy, the authors of Ref.~\cite{Kahlhoefer:2015vua} 
have obtained another lower
bound on $\sigma \over m$, ${\sigma \over m} \geqslant 1.5$ cm$^2$/g 
(more precisely its order lies $\sim 1.5 - 3.0$ cm$^2$/g). 
The analysis of the other astrophysical bodies such as
``bullet cluster" (1E 0657-56)~\cite{Randall:2007ph} sets the limit (upper) on the quantity $\sigma \over m$ to be
less than $\sim 1.25$ cm$^2$/g (68\% CL) considering the effect of drag force arising from 
DM particle collisions on the subcluster halo (particle momentum exchange is isotropic). 
A more stringent limit, ${\sigma \over m} < 0.7$ cm$^2$/g (68\% CL),
is also set by considering the effect of mass-to-light ratio of the 
subcluster due to the scattering of DM particles 
(particle momentum exchange is directional)~\cite{Randall:2007ph}.
Another tight constraint on self-interaction has been derived from the weak lensing analysis of
the observational data of the giant bullet cluster collision 1E0657-558~\cite{Clowe:2003tk} .
Also the studies on 72 clusters collisions data put an upper limit on
$\sigma \over m$, ${\sigma \over m} < 0.47$ cm$^2$/g with 95\% CL~\cite{Harvey:2015hha}. 
Clearly the obtained result of Ref.~\cite{Harvey:2015hha}
is in tension with the previous obtained bound from the observation of ``bullet cluster" (1E 0657-56).

Although various aspects of self-interacting DM scenario have widely been investigated 
in numerical simulation as well as in astrophysics, those for particle 
physics~\cite{Mambrini:2015nza} are comparatively less available in literature.
Also, studies on both DM velocity dependent and independent self-interactions have also been pursued.
While early studies mainly focus on the velocity-independent DM self interaction, there have been
recent developments in velocity-dependent scenario~\cite{Ackerman:2008gi, Feng:2009mn,
Feng:2009hw,Buckley:2009in,Ibe:2009mk,Loeb:2010gj,Tulin:2012wi,Tulin:2013teo} which can evade 
observed bounds by yielding different DM self scattering cross sections 
on various scales (dwarf galaxy, cluster etc.) and are thus interesting to study.
There are a variety of other proposed models for DM self interactions: 
mirror DM~\cite{Mohapatra:2001sx,Foot:2004wz,Foot:2012ai}, 
atomic DM~\cite{Kaplan:2011yj,CyrRacine:2012fz}, DM coupled to dark photon $\phi$ which
is originated hidden $U(1)^\prime$ symmetry~\cite{Ackerman:2008gi, Feng:2009mn,
Feng:2009hw,Buckley:2009in,Ibe:2009mk,Loeb:2010gj,Tulin:2012wi,Tulin:2013teo}.

Although there are a plethora of particle physics models for WIMP which can explain
various astrophysical as well as experimental observations, however, there are still lack of 
experimental or observational evidences to precisely distinguish the correct particle physics
model for DM physics. It may be possible to detect DM in future with direct or indirect detection
techniques but still these early observations may only provide the information about
the properties of DM particles but not the underlying theory. Therefore in order to avoid such
theoretical bias, model independent studies of DM phenomenology become particularly interesting
in recent studies~\cite{Birkedal:2004xn,Giuliani:2004uk,Kurylov:2003ra,Beltran:2008xg,
Cirelli:2008pk,Shepherd:2009sa}.

There are studies~\cite{Cao:2009uv,Cao:2009uw,Beltran:2010ww,Fitzpatrick:2010em,Goodman:2010yf,Bai:2010hh,
Goodman:2010ku,Goodman:2010qn,Bell:2010ei,Zheng:2010js,Yu:2011by} from the aspects of such model independent ways on DM phenomenology to relate 
the observed DM relic density, direct \& indirect detections and collider signatures.
In this paper we also consider a model independent WIMPs rather than choosing any particular
model. We follow the approach of effective 4-fermion interactions among the WIMPs for various 
combinations of spins and interaction forms~\cite{Beltran:2008xg, Zheng:2010js}. 
For this study we have simply considered a Dirac fermionic WIMP and 
construct the general effective 4-fermion operators for DM-DM self-interactions.
A detailed investigation and analysis of parameter space is performed thereafter. 
Moreover, we also evaluate and interpret the constraints for the self-interaction
of WIMPs for each case of spin and interaction forms from cluster observational results.

The paper is organised as follows. In Sec.~\ref{sec2} the effective Lagrangians for DM self-interaction
in this framework are briefly discussed. The calculational results for effective DM self-interaction 
cross sections are mentioned in Sec.~\ref{sec3}. The following section (Sec.~\ref{sec4}) is devoted to a detailed 
discussion on the parameters of this effective DM self-interaction scenario.
The effective self-couplings are constrained for a wide range of DM masses from several 
observational bounds on galaxy cluster scales in Sec.~\ref{sec5}. In Sec.~\ref{sec6} we finally summarise our work 
and some interesting conclusions are drawn.

\section{Effective Lagrangians for Dark Matter Self-Interaction \label{sec2}}

We consider the case of a single Dirac fermionic WIMP, $\chi$ for pursuing this analysis. In order to begin
our study we take effective operators between DM particles without considering any particular
preferred model for DM. However, we make some assumptions at this point. 
First of all, the WIMP is assumed to be the only new particle beyond SM 
at the electroweak scale. This implies that the calculations performed hereafter 
remain unaffected by possible other effects
(resonances etc.) and self interactions can thus safely be described by effective field
theoretical framework. Secondly we also assume, for simplicity, that the considered 4-fermion interactions 
among the DM fermions is governed by only one form of interaction (scalar, vector, tensor, etc.) 
among the possible set of effective operators.
Therefore the possible effective Lagrangians involving 
effective self interaction operators among four such WIMPs ($\chi$) can be given by,
\begin{eqnarray}
\text{Scalar (S)}:&&\quad \mathcal{L}_{\mathrm{S}}=G_{S}\,\bar{\chi}\chi\bar{\chi}\chi\,,\label{Leff-S}\\ [0.1cm]
\text{Pseudoscalar (P)}:&&\quad \mathcal{L}_{\mathrm{P}}=G_{P}\bar{\chi}\gamma_5\chi\bar{\chi}\gamma_5\chi\,,
\label{Leff-P}\\ [0.1cm]
\text{Vector (V)}:&&\quad \mathcal{L}_{\mathrm{V}}=G_{V}\,\bar{\chi}\gamma^\mu\chi\bar{\chi}\gamma_\mu \chi\,,\label{Leff-V}\\ [0.1cm]
\text{Axialvector (A)}:&&\quad \mathcal{L}_{\mathrm{A}}=G_{A}\,\bar{\chi}\gamma^\mu \gamma_5\chi\bar{\chi}\gamma_\mu\gamma_5 \chi\,,\label{Leff-A}\\ [0.1cm]
\text{Tensor (T)}:&&\quad \mathcal{L}_{\mathrm{T}}=G_{T}\bar{\chi}\sigma^{\mu\nu}\chi\bar{\chi}\sigma_{\mu\nu} \chi\,,\label{Leff-T}\\ [0.1cm]
\text{Scalar-pseudoscalar (SP)}:&&\quad \mathcal{L}_{\mathrm{SP}}=G_{SP}\,\bar{\chi}\chi\bar{\chi}i\gamma_5\chi\,,
\label{Leff-SP}\\ [0.1cm]
\text{Pseudoscalar-scalar (PS)}:&&\quad \mathcal{L}_{\mathrm{PS}}=G_{PS}\,\bar{\chi}i\gamma_5\chi\bar{\chi}\chi\,,
\label{Leff-PS}\\ [0.1cm]
\text{Vector-axialvector (VA)}:&&\quad \mathcal{L}_{\mathrm{VA}}
=G_{VA}\,\bar{\chi}\gamma^\mu\chi\bar{\chi}\gamma_\mu\gamma_5\chi\,,
\label{Leff-VA}\\ [0.1cm]
\text{Axialvector-vector (AV)}:&&\quad \mathcal{L}_{\mathrm{AV}}
=G_{AV}\,\bar{\chi}\gamma^\mu\gamma_5\chi\bar{\chi}\gamma_\mu \chi\,,\label{Leff-AV}\\ [0.1cm]
\text{Alternative tensor ($\tilde{\mathrm{T}}$)}:&&\quad \mathcal{L}_{\tilde{\mathrm{T}}}
=G_{\tilde{T}}\,\varepsilon^{\mu\nu\rho\sigma}
\bar{\chi}\sigma_{\mu\nu}\chi\bar{\chi}\sigma_{\rho\sigma} \chi\,,\label{Leff-T-tild}\\ [0.1cm]
\text{Left handed-left handed (LL)}:&&\quad \mathcal{L}_{\mathrm{LL}}
=G_{LL}\,
\bar{\chi}\gamma^\mu(1-\gamma_5)\chi\bar{\chi}\gamma_\mu(1-\gamma_5)\chi\,,\label{Leff-LL}\\ [0.1cm]
\text{Right handed-right handed (RR)}:&&\quad \mathcal{L}_{\mathrm{RR}}
=G_{RR}\,
\bar{\chi}\gamma^\mu(1+\gamma_5)\chi\bar{\chi}\gamma_\mu(1+\gamma_5)\chi\,,\label{Leff-RR}\\ [0.1cm]
\text{Left handed-right handed (LR)}:&&\quad \mathcal{L}_{\mathrm{LR}}
=G_{LR}\,
\bar{\chi}\gamma^\mu(1-\gamma_5)\chi\bar{\chi}\gamma_\mu(1+\gamma_5)\chi\,,\label{Leff-LR}\\ [0.1cm]
\text{Right handed-left handed (RL)}:&&\quad \mathcal{L}_{\mathrm{RL}}
=G_{RL}\,
\bar{\chi}\gamma^\mu(1+\gamma_5)\chi\bar{\chi}\gamma_\mu(1-\gamma_5)\chi\,,\label{Leff-RL}
\end{eqnarray}
where the effective coupling constants $G$s are the real number with mass dimension $-2$. In the
above $\gamma_\mu, \gamma_5$ are the usual gamma matrices and $\sigma_{\mu\nu}, 
\varepsilon^{\mu\nu\rho\sigma}$ are the antisymmetric rank-2 tensor and Levi-Civita symbol
respectively. 
We tag the above interaction types based on the $2\rightarrow 2$ self-scattering processes of 
fermionic DM particles, namely, $\bar{\chi} \chi \to \bar{\chi} \chi$ using effective theoretical
approach. It should be noted that the considered effective operators shown in 
Eqs.~\ref{Leff-S}-\ref{Leff-RL} are invariant under both CPT and Lorentz transformations.
Also, the first five operators in Eqs.~\ref{Leff-S}-\ref{Leff-T} are separately invariant under
C, P and T transformations.
In addition, they also respect the Hermiticity requirements.
All of the above listed operators do not guarantee independent self interaction cross sections.
For example, the effective Lagrangian for scalar-pseudoscalar (SP) interaction, $\mathcal{L}_{\mathrm{SP}}$ 
of Eq.~\ref{Leff-SP} and that for pseudoscalar-scalar (PS) interaction, $\mathcal{L}_{\mathrm{PS}}$ 
of Eq.~\ref{Leff-PS} are similar
in nature for the case of effective 4-point diagrams of DM self-interaction and can yield similar results.
Similarly, the effective Lagrangian for vector-axialvector (VA) operator of Eq.~\ref{Leff-VA} and that for 
axialvector-vector operator of Eq.~\ref{Leff-AV} result equal squared transition amplitude 
(or squared scattering amplitude) for effective self-interacting diagrams. 
Moreover, the last four effective Lagrangians 
of Eqs.~\ref{Leff-LL}, \ref{Leff-RR}, \ref{Leff-LR} and \ref{Leff-RL}, namely, 
$\mathcal{L}_{\mathrm{LL}}$, $\mathcal{L}_{\mathrm{RR}}$, $\mathcal{L}_{\mathrm{LR}}$ and 
$\mathcal{L}_{\mathrm{RL}}$ respectively, represent chiral (C) interactions among DM particles and 
also produce squared amplitude terms for DM self-scattering diagrams similar to each other.
Hence from the above listed set of operators we are left with nine operators 
that can result in producing 
independent self scattering cross sections in case of Dirac fermionic DM.
Note that the chiral interaction operators can be derived from suitable combinations of 
other listed operators. Also there may be some other forms for alternative tensor ($\tilde{\rm T}$)
in Eq.~\ref{Leff-T-tild} such as $\bar{\chi}\sigma^{\mu\nu}i\gamma_5\chi\bar{\chi}\sigma_{\mu\nu}\chi$,
$\bar{\chi}\sigma^{\mu\nu}\chi\bar{\chi}\sigma_{\mu\nu}i\gamma_5 \chi$ which are actually equivalent
to $\varepsilon^{\mu\nu\rho\sigma}\bar{\chi}\sigma_{\mu\nu}\chi\bar{\chi}\sigma_{\rho\sigma} \chi$.

\begin{figure}[tbc!]
\begin{center}
\includegraphics[scale=0.7,angle=0]{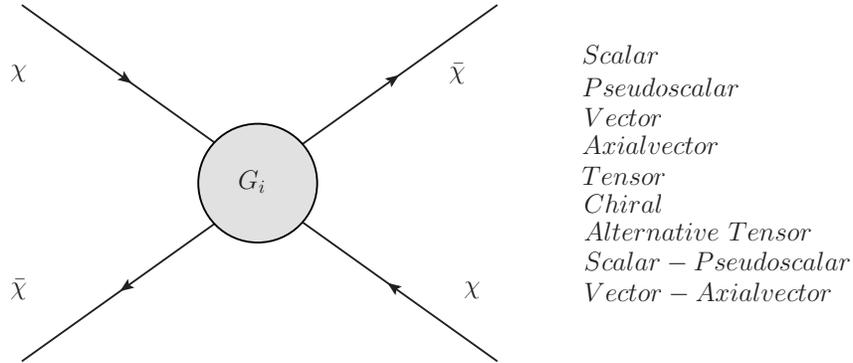}
\caption{ {\it Feynman diagram for self interaction of fermionic dark matter ($\chi$) in effective theory.
See text for more details.}}
   \label{sidm}
\end{center}
\end{figure}

The basic Feynman diagram for the aforementioned effective operators is depicted in Fig.~\ref{sidm}.

\section{Effective Dark Matter Self Interaction Cross Sections \label{sec3}}

In order to continue our analysis we calculate the DM self scattering cross-sections 
for self interaction considering the above mentioned operators and get the following
\begin{eqnarray}
\sigma_{S,\,\mathrm{self}}&=&\frac{1}{16\pi } G_{S}^2 \,\,
\frac{{(s-4m_\chi^2)}^2}{s} \, ,\label{diracWIMP-sigma-S}\\
\sigma_{P,\,\mathrm{self}}&=&\frac{1}{16\pi} G_{P}^2 \,\, s \, ,
\label{diracWIMP-sigma-P}\\
\sigma_{V,\,\mathrm{self}}&=&\frac{1}{12\pi}
G_{V}^2\,\,
\bigg[s+4m_\chi^2+4\frac{m_\chi^4}{s}\bigg] \, ,\label{diracWIMP-sigma-V}\\
\sigma_{A,\,\mathrm{self}}&=&\frac{1}{12\pi}
G_{A}^2\,\,
\bigg[s-8m_\chi^2+28\frac{m_\chi^4}{s}\bigg] \, ,\label{diracWIMP-sigma-A}\\
\sigma_{T,\,\mathrm{self}}&=&\frac{1}{6\pi}
G_{T}^2 \,\,
\bigg[s+4m_\chi^2+40\frac{m_\chi^4}{s}\bigg] \, ,\label{diracWIMP-sigma-T}\\[0.1cm]
\sigma_{K,\,\mathrm{self}}&\equiv&\sigma_{SP,\,\mathrm{self}}=\sigma_{PS,\,\mathrm{self}}
=\frac{1}{16\pi}
G_{K}^2 \,\,
(s-4m_\chi^2) \, ,\label{diracWIMP-sigma-SP-PS}\\[0.3cm]
\sigma_{M,\,\mathrm{self}}&\equiv&\sigma_{VA,\,\mathrm{self}}=\sigma_{AV,\,\mathrm{self}}
=\frac{1}{12\pi}
G_{M}^2 \,\,
\bigg[s-2m_\chi^2-8\frac{m_\chi^4}{s}\bigg] \, ,\label{diracWIMP-sigma-AV-VA}\\[0.3cm]
\sigma_{\tilde{T},\,\mathrm{self}}&=&\frac{2}{3\pi}
G_{\tilde{T}}^2 \,\,
\bigg[s+4m_\chi^2-32\frac{m_\chi^4}{s}\bigg] \, ,\label{diracWIMP-sigma-T-tild}\\[0.3cm]
\sigma_{C,\,\mathrm{self}}&\equiv&\sigma_{LL,\,\mathrm{self}}=\sigma_{RR,\,\mathrm{self}}
=\sigma_{LR,\,\mathrm{self}}=\sigma_{RL,\,\mathrm{self}}\nonumber\\
&=&\frac{1}{3\pi} G_{C}^2 \,\,
\bigg[s-2m_\chi^2+4\frac{m_\chi^4}{s}\bigg] \, ,\label{diracWIMP-sigma-chiral}
\end{eqnarray}
where $s$ is the usual Mandelstem variable, $m_{\chi}$ is the mass of fermionic WIMP $\chi$. Now since
in the non-relativistic limit the relative velocity can be written in terms of mass of the WIMP
($\chi$) and the Mandelstem variable $s$ as,
\bea
v_{rel} = {\sqrt{s(s-4m^2_{\chi})}\over (s-2m^2_{\chi})}\,\,\, .
\eea
Hence in the lab frame $s$ can be rewritten in terms of the relative velocity, $v_{\rm rel}$ as
\begin{flalign}
s=2m^2_{\chi} \left(1+\frac{1}{\sqrt{1-v^2_\text{rel}}}\right).
\label{s_vrel}
\end{flalign}
Also in this context it should be mentioned that if the velocity of two colliding DM particles 
are $\vec{v}_1$ and $\vec{v}_2$, then $v_{\rm rel}$ can be
written as
\begin{flalign}
v_{\rm rel}=|\vec{v}_{1}-\vec{v}_{2}|,
\label{v_r_def}
\end{flalign}
in the non-relativistic limit, 
while in the case of relativistic velocities, 
the relative velocity between such DM particles is given by~\cite{Cannoni:2015wba} 
\begin{flalign}
v_{\text{rel}}=
\frac{
\sqrt{(\vec{v}_1 - \vec{v}_2)^2 - 
\frac{(\vec{v}_1 \times \vec{v}_2)^2}{c^2}
}}
{1-\frac{\vec{v}_1 \cdot \vec{v}_2}{c^2}}\, .
\label{v_rel_Rel_def}
\end{flalign} 
Velocity of light $c$ is chosen to be unity and hence $v_{rel}$ is dimensionless 
in our later calculations.

Substituting the expression of $s$ of Eq.~\ref{s_vrel}, 
Eqs.~\ref{diracWIMP-sigma-S}-\ref{diracWIMP-sigma-chiral} can be recasted as
\begin{eqnarray}
\sigma_{S,\,\mathrm{self}}&=&\frac{1}{16\pi } G_{S}^2\,\,
2m^2_{\chi}\left[\frac{{\left({1\over\sqrt{1-v^2_\text{rel}}} - 1\right)}^2}{1 + {1\over\sqrt{1-v^2_\text{rel}}}}\right] \, ,\label{sigmaS}\\
\sigma_{P,\,\mathrm{self}}&=&\frac{1}{16\pi}
G_{P}^2\,\,
2m^2_{\chi}\bigg[1+{1\over\sqrt{1-v^2_\text{rel}}}\bigg] \, ,
\label{sigmaP}\\
\sigma_{V,\,\mathrm{self}}&=&\frac{1}{12\pi}
G_{V}^2\,\,
2m^2_{\chi}\bigg[3+{1\over\sqrt{1-v^2_\text{rel}}}+{1\over1+{1\over\sqrt{1-v^2_\text{rel}}}}\bigg] \, ,\label{sigmaV}\\
\sigma_{A,\,\mathrm{self}}&=&\frac{1}{12\pi}
G_{A}^2 \,\,
2m^2_{\chi}\bigg[-3+{1\over\sqrt{1-v^2_\text{rel}}}+{7\over1+{1\over\sqrt{1-v^2_\text{rel}}}}\bigg] \, ,\label{sigmaA}\\
\sigma_{T,\,\mathrm{self}}&=&\frac{1}{6\pi}
G_{T}^2\,\,
2m^2_{\chi}\bigg[3+{1\over\sqrt{1-v^2_\text{rel}}}+{10\over1+{1\over\sqrt{1-v^2_\text{rel}}}}\bigg] \, ,\label{sigmaT}\\[0.1cm]
\sigma_{K,\,\mathrm{self}}&\equiv&\sigma_{SP,\,\mathrm{self}}=\sigma_{PS,\,\mathrm{self}}
=\frac{1}{16\pi}
G_{K}^2\,\,
2m^2_{\chi}\bigg[-1+{1\over\sqrt{1-v^2_\text{rel}}}\bigg] \, ,\label{sigmaK}\\[0.2cm]
\sigma_{M,\,\mathrm{self}}&\equiv&\sigma_{VA,\,\mathrm{self}}=\sigma_{AV,\,\mathrm{self}}
=\frac{1}{12\pi}
G_{M}^2\,\,
2m^2_{\chi}\bigg[{1\over2}+{1\over\sqrt{1-v^2_\text{rel}}}-{2\over1+{1\over\sqrt{1-v^2_\text{rel}}}}\bigg] \, ,\label{sigmaM}\\[0.2cm]
\sigma_{\tilde{T},\,\mathrm{self}}&=&\frac{2}{3\pi}
G_{\tilde{T}}^2\,\,
2m^2_{\chi}\bigg[3+{1\over\sqrt{1-v^2_\text{rel}}}-{8\over1+{1\over\sqrt{1-v^2_\text{rel}}}}\bigg]\, ,\label{sigmaT-tild}\\[0.2cm]
\sigma_{C,\,\mathrm{self}}&\equiv&\sigma_{LL,\,\mathrm{self}}=\sigma_{RR,\,\mathrm{self}}
=\sigma_{LR,\,\mathrm{self}}=\sigma_{RL,\,\mathrm{self}}\nonumber\\
&=&\frac{1}{3\pi} G_{C}^2\,\,
2m^2_{\chi}\bigg[{1\over\sqrt{1-v^2_\text{rel}}}+{1\over1+{1\over\sqrt{1-v^2_\text{rel}}}}\bigg] .\label{sigmaC}
\end{eqnarray}
In the above, since $\sigma_{SP,\,\mathrm{self}}$ and $\sigma_{PS,\,\mathrm{self}}$ yield similar 
results, these two terms are denoted by $\sigma_{K,\,\mathrm{self}}$ (say). Similarly,
$\sigma_{M,\,\mathrm{self}}$ represents the terms $\sigma_{VA,\,\mathrm{self}}$ and 
$\sigma_{AV,\,\mathrm{self}}$ while $\sigma_{C,\,\mathrm{self}}$ is for the chiral (LL, RR, LR, RL) ones.
Now, it can be noted from Eqs.~\ref{sigmaS}-\ref{sigmaC} that the DM self scattering cross sections
depend only on three parameters, namely, the effective couplings, DM relative velocity and DM mass.

\section{Parametric Dependence of Dark Matter Effective Self Interaction Cross Sections \label{sec4}}

In order to inspect the general nature of Eqs.~\ref{sigmaS}-\ref{sigmaC}, 
all of the above equations (Eqs.~\ref{sigmaS}-\ref{sigmaC}) can be recasted in terms of 
various types of scattering cross-sections ($\sigma_i$) in a general form as,
\bea
\sigma_i = C_i\, G^2_i\, 2m ^2_{\chi}\, f_i (v_{\rm rel}) \, ,
\label{gen_si}
\eea
where $G_i$s are the effective couplings and $f_i (v_{\rm rel})$ are different functions
of DM relative velocity for different operator types. In the above $C_i$ are constants
which depend on phase-space factors for different types of interactions mentioned before. 
Now it can be seen from Eq.~\ref{gen_si} that for any type of effective self-interacting DM
scenario, the self-scattering cross-section is proportional to the square of both effective
self-coupling and DM mass, i.e., $G^2_i m ^2_{\chi}$. 
Thus for any type of self-interaction in such effective interaction theory, 
$\sigma \propto G^2 m^2_{\chi}$
for a particular DM relative velocity.
Since from various astrophysical as well as
cosmological observations one can put limits on the factor, $\sigma \over m_{\chi}$, this could
restrict the limit on the parameter space containing effective self-coupling, DM mass and its 
relative velocity. In this context it is worth important to note that the DM relative velocity
depends on the thermal distribution of DM particles. If the DM relative velocity is estimated
from such thermal distribution of DM, one can, in principle, estimate the limits on the 
effective self-coupling for different DM masses from the precisely measured DM self-scattering
cross-section. 

\begin{figure}[tbc!]
\begin{center}
\includegraphics[scale=0.7,angle=-90]{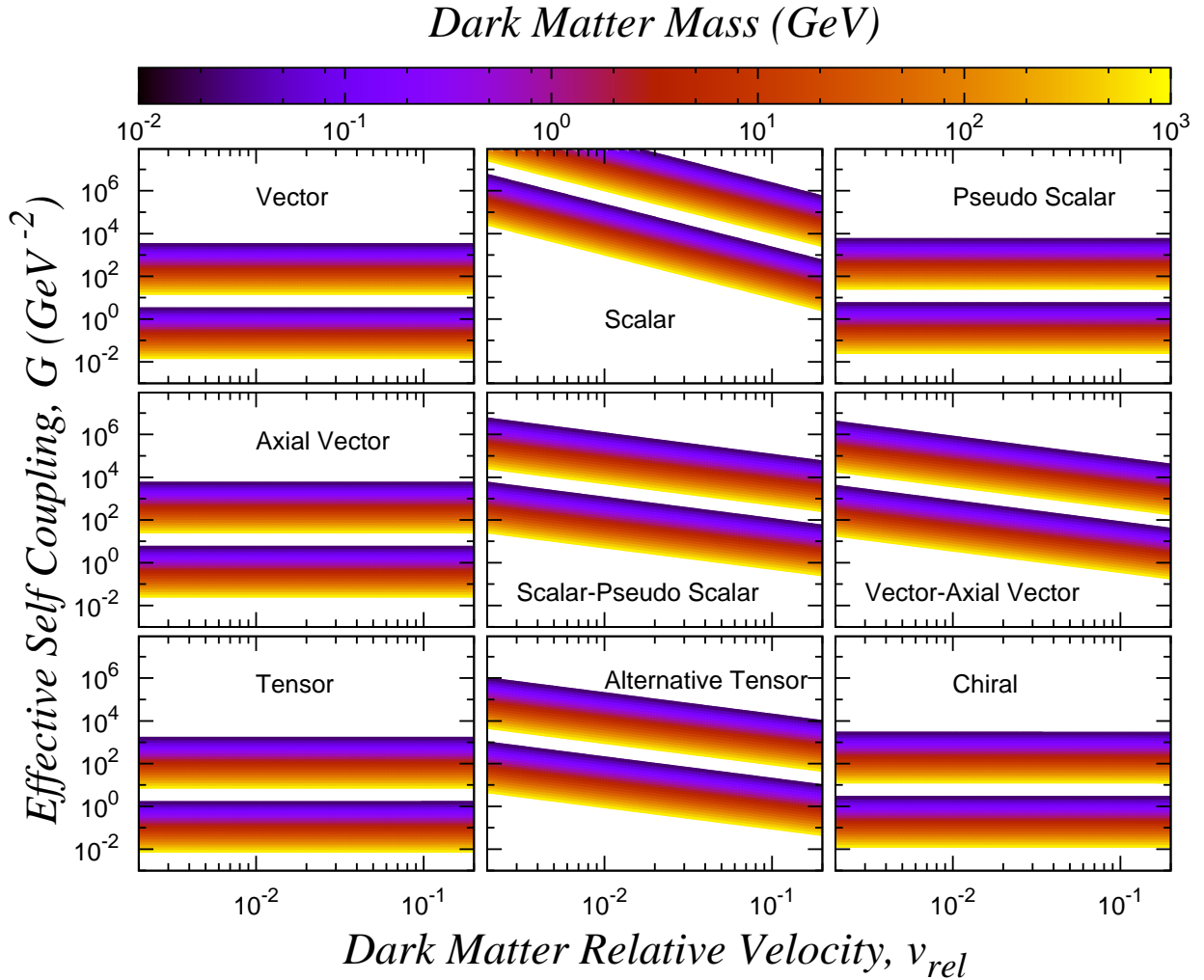}
\caption{ {\it Variations of effective self couplings ($G_{i}$) with both dark matter relative 
velocity ($v_{\rm rel}$) and dark matter mass ($m_{\chi}$) for different possible 
types of self-interactions discussed in the text. The colour gradient indicates the 
variation of dark matter mass. The blueish coloured region in the colour bar denotes 
the lower mass zone while the yellowish one is for the higher mass. The two band-like
regions in each plot are for two considered values of self-interaction cross-sections,
namely, for ${\sigma \over m} = 5\times10^4$ GeV$^{-3}$ (10.0 cm$^2$/g) and $5\times10^{-2}$ 
GeV$^{-3}$ ($1.0\times10^{-5}$ cm$^2$/g). 
See text for more details.}}
   \label{self_all}
\end{center}
\end{figure}

In order to elaborate how the self-interaction parameters depend on each other in this
effective theory, we make a 3-D colour coded plot in Fig.~\ref{self_all} where
the variations of effective self couplings ($G_{i}$) with the simultaneous 
variations of other parameters, namely, dark matter relative
velocity ($v_{\rm rel}$) and dark matter mass ($m_{\chi}$) are furnished.  
In  Fig.~\ref{self_all} we have estimated such variations for the above-mentioned 
nine independent operators of self-interaction processes, namely, for scalar (S), vector (V), pseudoscalar (P),
axialvector (A), scalar-pseudoscalar (SP), vector-axialvector (VA), tensor (T), 
alternative tensor ($\tilde{\rm T}$) and chiral (C)
interactions. The colour index in Fig.~\ref{self_all} denotes the DM mass where the DM mass
varies from bluish coloured region to yellowish region as it increases.
For all of the plots in Fig.~\ref{self_all} the mass of DM particles has been varied
from 10 MeV to 1 TeV as shown by the colour coded index. However, we can, in principle,
choose any other mass ranges of DM but the resulting plots will be similar in nature.  
Moreover, we have considered two different values of self-interaction cross-sections, namely,
${\sigma \over m} = 5\times10^4$ GeV$^{-3}$ ($10$ cm$^2$/g) and 
$5\times10^{-2}$ GeV$^{-3}$ ($1.0\times10^{-5}$ cm$^2$/g) for each of these plots.
\footnote{Note that 1 GeV$^{-3}$ = $2\times 10^{-4}$ cm$^2$/g, 1 barn/GeV = 0.6 cm$^2$/g}
This is because of the fact that the self-interaction is not precisely measured
since and the measurements of the factor, $\sigma \over m_{\chi}$ from various observations
vary much significantly (few orders of magnitude) due to various astrophysical uncertainties.
Hence we restrict the values of the factor, $\sigma \over m_{\chi}$ to those which may be
the predicted extreme limits on self interaction from various studies~\cite{Mambrini:2015nza}.
We can, in principle, choose any other limiting values of $\sigma \over m_{\chi}$ which
were estimated to confront several astrophysical observations such  
as bullet cluster etc. This would rather change the limits on the effective self-couplings.
But the main motivation for this plot remains unchanged.
The two band-like structures in each frame of Fig.~\ref{self_all} are due to such two choices
of ${\sigma \over m}$. In each plot the upper band corresponds to the chosen 
higher value of ${\sigma \over m}$ (10.0 cm$^2$/g or $5\times10^4$ GeV$^{-3}$) 
while the band on the lower side 
is for the lower value ($1.0\times10^{-5}$ cm$^2$/g or $5\times10^{-2}$ GeV$^{-3}$).
In Fig.~\ref{self_all} we see that for a particular DM mass the effective self-couplings
remain almost constant with a broad range of DM relative velocities for the
self-couplings to be vector (V), axialvector (A), pseudoscalar (P), tensor (T) and
chiral (C) in nature. On the other hand, for the cases of effective self-couplings due to 
scalar (S), scalar-pseudoscalar (SP), vector-axialvector (VA) and 
alternative tensor ($\tilde{\rm T}$) interactions, 
these effective self-couplings decrease with increasing DM relative velocities
as shown in Fig.~\ref{self_all}. 
However, at a very large DM velocities, i.e., for the case of
relativistic self-collisions of DM particles, the effective self-couplings decrease
very steeply with DM relative velocity for all types of self-couplings.
It is clear from Fig.~\ref{self_all} that 
these above-mentioned variations of different self-couplings with DM relative velocity are 
similar for any chosen values of DM mass and self-interaction cross-section.
It is also obvious from this figure that for a particular values of $v_{\rm rel}$ and 
$\sigma \over m_{\chi}$, if the DM mass becomes very high 
(towards yellowish coloured zone in Fig.~\ref{self_all}),
the value of self-coupling decreases compared to the case where DM mass is smaller
(towards bluish coloured zone in Fig.~\ref{self_all}).

\begin{figure}[tbc!]
\begin{center}
\includegraphics[scale=0.6,angle=-90]{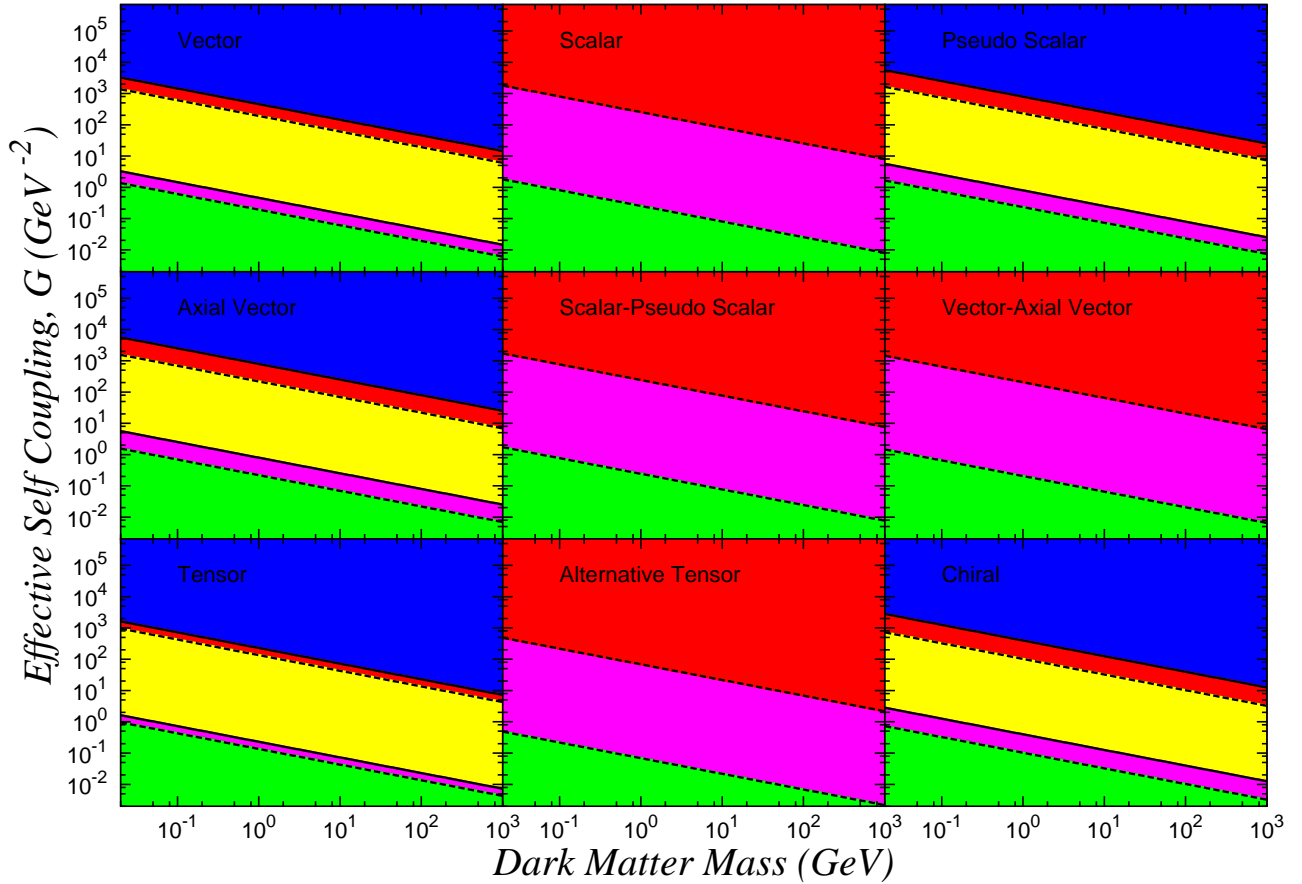}
   \caption{ {\it Plot showing the allowed parameter space in the plane containing 
   effective self couplings ($G_{i}$) and fermionic dark matter mass ($m_{\chi}$) for different possible
   self interaction types. The black dotted line in each plot denotes dark matter 
   relative velocity, $v_{\rm rel} = 0.999c$ (relativistic collision) while the solid black 
   line is for $v_{rel} = 0$ (non-relativistic collision).
   The red and magenta coloured regions in each plot is for two different self interaction
   cross section values, namely, 
   for ${\sigma \over m} = 5\times10^4$ GeV$^{-3}$ ($10.0$ cm$^2$/g) and 
   $5\times10^{-2}$ GeV$^{-3}$ ($1.0\times10^{-5}$ cm$^2$/g) respectively.
   The green and blue region denote the allowed and excluded parameter space for the above 
   two values of ${\sigma \over m}$ whereas the yellow region is allowed for 
   ${\sigma \over m} = 0.05$ GeV$^{-3}$ but excluded for the other ${\sigma \over m}$ value.
   See text for more details.}}
   \label{si_all}
\end{center}
\end{figure}

We finally estimate the dependences of various types of effective self-couplings ($G_{i}$)
on different mass scales of fermionic DM ($m_{\chi}$). 
In order to compute such dependencies we again only consider
the two recently reported values of DM self-interaction on Abell 3827 observation.
Needless to mention here that we can, in principle, consider any reported DM self-interaction
values from astrophysical or cosmological observations as well as from numerical simulations 
other than those chosen in this work. But this may change the constraints on 
the effective self-couplings from that obtained in this work. Since there are 
conflicts among different reported 
self-interaction values, we keep our options open in this work. 
We show the allowed zone of all possible types of effective self couplings ($G_{i}$) for 
different dark matter masses ($m_{\chi}$) in Fig.~\ref{si_all}. 
We would like to mention at this point that similar to Fig.~\ref{self_all}, 
here also the considered interactions include scalar (S), vector (V), pseudoscalar (P),
axialvector (A), scalar-pseudoscalar (SP), vector-axialvector (VA), tensor (T), 
alternative tensor ($\tilde{\rm T}$) and chiral (C) types.
In all of these
plots we have taken into consideration the two extreme values of DM relative velocity ($v_{\rm rel}$),
namely, $v_{rel} = 0.999c$ and 0 ($c$ is velocity of light in vacuum). 
These velocities correspond to the relativistic and 
non-relativistic regimes of DM self-scattering respectively. 
In Fig.~\ref{si_all} the black dotted lines indicate dark matter 
relative velocity, $v_{rel} = 0.999c$ while the black solid lines are for the other one, $v_{rel} = 0$.
The red and magenta coloured regions in each plot of Fig.~\ref{si_all} denote the allowed zone for two 
considered self-interaction cross section values, namely, 
for ${\sigma \over m} = 5\times10^4$ GeV$^{-3}$ and $5\times10^{-2}$ GeV$^{-3}$ respectively.
Also in each plot of Fig.~\ref{si_all} the green and blue regions denote the parameter space 
for values of ${\sigma \over m} < 1.0\times10^{-5}$ cm$^2$/g ($5\times10^{-2}$ GeV$^{-3}$) 
and ${\sigma \over m} > 10$ cm$^2$/g ($5\times10^{4}$ GeV$^{-3}$) respectively,
whereas the yellow region is allowed for the values of DM self-interaction in between the chosen ones,  
$1.0\times10^{-5}$ cm$^2$/g ($5\times10^{-2}$ GeV$^{-3}$) $< {\sigma \over m} < 10$ cm$^2$/g ($5\times10^{4}$ GeV$^{-3}$).
We see in Fig.~\ref{si_all} that the allowed zone of effective self couplings, $G_{i}$ 
decrease with increasing DM mass. This simply corresponds to the fact that self-interaction of
more massive fermionic DM particles happens through a heavier messenger
particle while a lighter messenger is responsible for the
self-interaction of less massive fermionic DM.
The black solid and dotted lines in each plot of Fig~\ref{si_all} gives the upper and lower bounds on
$G_{i}$ for different DM masses, $m_{\chi}$ for a particular value of self-interaction. 
Since we have considered two values of ${\sigma \over m}$ for our calculations, 
there should be two black solid and dotted lines in each plots of Fig.~\ref{si_all}. 
But in Fig.~\ref{si_all} it can be seen that there in no black solid line ($v_{\rm rel} = 0$) appearing in
few frames, namely, for the cases of scalar (S), scalar-pseudoscalar (SP), vector-axialvector (VA)
and alternative tensor ($\tilde{\rm T}$) interactions. So, there is no restriction on upper bounds for
those interactions. This is due to the nature of calculated theoretical forms of $f_i (v_{\rm rel})$ 
in self-annihilation cross sections for those types of interactions. Therefore, the constraints on such types of
self-interaction is relaxed to higher values if DM particles interact much more non-relativistically.
Hence the blue regions are shifted to much higher values of $G_{i}$ and do not appear in those plots. 
Also since the magenta and red regions get partially overlapped 
with each other due to such shifts of black solid lines in those plots, the yellow regions extend 
up to large values and get overlapped by these magenta and red regions. Thus only lower bounds on
$G_{i}$ for different $m_{\chi}$ exist in those cases, which are denoted by black dotted lines.  

\section{Constraining Effective Dark Matter Self Couplings from Observations in Cluster Scales \label{sec5}}

In the realistic cases, the DM particles do not traverse in space with a fixed velocity but follow 
a velocity distribution such as Maxwell-Boltzmann velocity distribution. Therefore
the DM scattering probability in the DM halo is not fixed by only a single value of DM relative 
velocity, $v_{\rm rel}$ but by the convolution over different DM densities and velocities, 
precisely determined by N-body simulations. Here for simplicity we make an average of 
the scattering cross section over the velocity distributions of interacting DM particles. 
This velocity-averaged self-interacting cross section of DM ($\langle \sigma \rangle$) 
can be approximately treated as the measured quantity on which various astrophysical
observations put different constraints~\cite{Tulin:2013teo}. 
Since the WIMPs are assumed to be thermally distributed
we consider Maxwellian velocity distribution for the DM particles of the form
\bea
 \mathcal{F}(v) = {1\over (\pi v_0^2)^{3\over2}} \, e^{-v^2/v_0^2}\, ,
\eea
where $v_0$ is the most probable speed of DM. 
The choice of $v_0$ depends on the different characteristic
velocity scales of different halo sizes. 
If the initial DM velocities are $\vec{v}_1$ and $\vec{v}_2$, this averaging over
DM velocity distributions of such DM particles gives,
\bea
\langle \sigma \rangle &=& \int d^3 v_1 d^3 v_2 \, \mathcal{F}(v_1) \, \mathcal{F}(v_2) \, 
\sigma(|\vec{v}_1 -\vec{v}_2|) \nonumber \\ 
&=& \int \frac{ d^3 v_1 d^3 v_2}{(\pi v_0^2)^{3}} \, e^{-v_1^2/v_0^2}  \, e^{-v_2^2/v_0^2} \, 
\sigma(|\vec{v}_1 -\vec{v}_2|) \nonumber \\
&=& \int \frac{d^3 v_{\rm rel}}{(2\pi v_0^2)^{3/2}} \, e^{-\frac{1}{2} v_{\rm rel}^2/v_0^2} \, \sigma(v_{\rm rel}) \; ,
\eea
where $v_{\rm rel} = |\vec{v}_1 - \vec{v}_2|$ is the DM relative velocity as discussed earlier. 
The characteristic DM velocity scale in galaxy clusters is approximately $\sim 1000$ km/s whereas
for Milky Way galaxy such characteristic scale becomes $\sim 100$ km/s. On the other hand, in 
case of dwarf galaxy the characteristic velocity takes the value of $\sim 10$ km/s,
comparatively much smaller than the aforementioned velocity scales of other systems. 
We choose the value of $v_{rel}$ to be $\sim 1000$ km/s or $\sim 3.3\times 10^{-3}$ 
(in the unit of velocity of light in vacuum, $c\approx3\times10^5$ km/s) 
since we consider the reported observational bounds on DM self-interaction
in cluster scales. To proceed we compute the velocity-averaging of different DM self-interaction 
cross sections given in Eqs.~\ref{sigmaS}-\ref{sigmaC}. The computed values of velocity-averaging
of functions such as ${1\over\sqrt{1-v^2_\text{rel}}}$ and ${1\over1+{1\over\sqrt{1-v^2_\text{rel}}}}$,
appearing multiple times in different forms of $f_i (v_{\rm rel})$, come out to be 
$\sim$ 1.00002 and $\sim$ 0.5 respectively. In this method the values of $\langle \sigma \rangle$ for 
different types of aforementioned self interactions are derived for a typical cluster velocity 
scale of $\sim$ 1000 km/s. The computed ratio of $\langle \sigma \rangle$ to DM mass 
(${\langle \sigma \rangle \over m}$) in this effective framework may be safely treated as 
the observed constrained quantity (${\sigma \over m}\big|_{obs}$) for self interaction in cluster scales.
Therefore the calculated quantity ${\langle \sigma \rangle \over m}$ may be used to put constraints 
on the parameter space of this framework containing effective self-couplings ($G_{i}$) 
and DM mass ($m_{\chi}$).

Needless to mention at this point that the above-mentioned velocity-averaging is not
relevant if the cross section is independent of DM velocity whereas this averaging becomes
more important in case of strongly velocity-dependent cross sections 
(for example, resonance effects etc.).

\begin{figure}[tbc!]
\begin{center}
\includegraphics[scale=0.6,angle=-90]{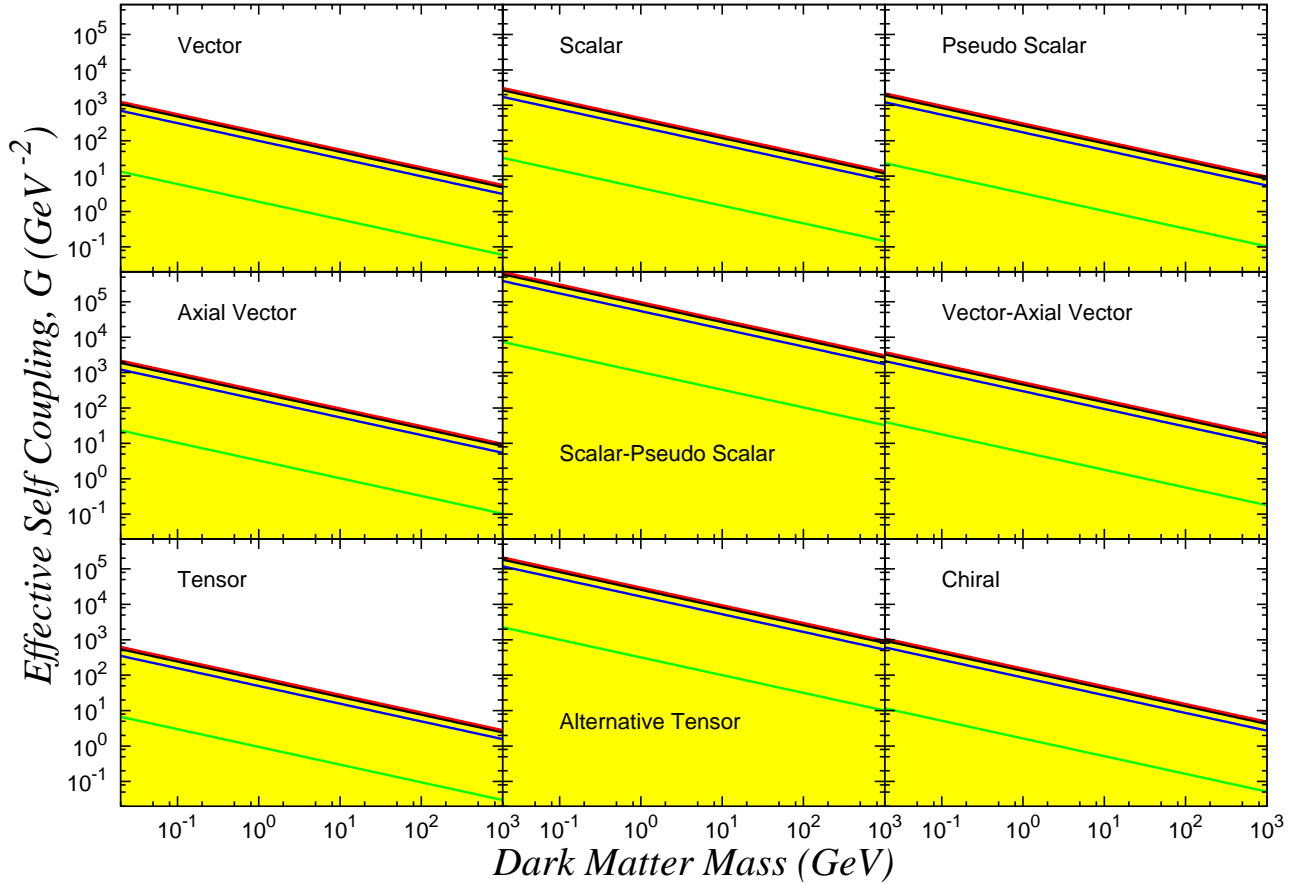}
\caption{ {\it Constraining various possible types of effective self couplings ($G_{i}$) for 
   different fermionic dark matter masses ($m_{\chi}$) from the observational results of 
   galaxy clusters. In each plot the solid green line denotes the reported limit on the
   the ratio beween dark matter self-interaction and mass, 
   ${\sigma \over m} = 1.7\times10^{-4}$ cm$^2$/g from the recent observational data of 
   Abell 3827 cluster while the solid red line is for another 
   claimed limit on ${\sigma \over m}$, namely, ${\sigma \over m} = 1.5$ cm$^2$/g 
   from the observation of same cluster. On the other hand, the upper limit on dark matter 
   self interaction (${\sigma \over m} < 0.47$ cm$^2$/g) from the recent study 
   of 72 galaxy clusters is shown by the solid blue line in each plot whereas 
   the black solid line denotes the conservative limit (upper) on such self interaction
   (${\sigma \over m} < 1.25$ cm$^2$/g) from the bullet cluster 1E 0657-56 observation.
   The yellow region represents the total allowed region of parameter space for 
   ${\sigma \over m} < \mathcal{O}(1)$ cm$^2$/g reported by various cluster observations. 
   See text for more details.
   }}
   \label{si_clust}
\end{center}
\end{figure}

In order to give a much more clear picture of the constrained parameter space, 
we have calculated the effective self couplings constrained by various observations 
of galaxy clusters for different DM mass ranges and shown the constraint parameter space 
containing the effective self couplings ($G_{i}$) and the DM mass ($m_{\chi}$) in Fig.~\ref{si_clust}. 
Clearly, each frame of Fig.~\ref{si_clust} represents a particular kind of 
DM self-interaction scenario discussed earlier.
In each plot of Fig.~\ref{si_clust} the solid green line represents 
${\sigma \over m} = 1.7\times10^{-4}$ cm$^2$/g from the recent observation (central value) of 
Abell 3827 cluster~\cite{Massey:2015dkw} while the solid red line denotes for the another 
claimed limit on ${\sigma \over m}$, namely, for ${\sigma \over m} = 1.5$ cm$^2$/g 
from the study of the same cluster~\cite{Kahlhoefer:2015vua}.
On the other hand, the upper limit on DM self interaction (${\sigma \over m} < 0.47$ cm$^2$/g) 
from the analysis of observational data of 72 galaxy clusters~\cite{Harvey:2015hha} is shown by the 
solid blue line in each plot whereas 
the black solid line is for the conservative upper limit on DM self interaction
(${\sigma \over m} < 1.25$ cm$^2$/g) from the observation of bullet cluster (1E 0657-56)~\cite{Randall:2007ph}.
The total allowed region of parameter space for
${\sigma \over m} < \mathcal{O}(1)$ cm$^2$/g as reported by various cluster observations
is shown by the yellow coloured region in each plot of Fig.~\ref{si_clust}.
Since different observational as well as simulation-based results in the recent past 
for the cluster scales impose
severe constraints on the quantity ${\sigma \over m}$ so that its numerical value 
cannot exceed $\sim \mathcal{O}(1)$ cm$^2$/g, the parameter space outside these yellow regions
in Fig.~\ref{si_clust} is also not viable for those results. This also implies that for any limit 
of ${\sigma \over m}$ to be less than $\mathcal{O}(1)$ cm$^2$/g, the projected values of effective 
self couplings lie on the yellow regimes for the range of DM masses shown in this figure.
Therefore any projected limit on
DM self-interaction for cluster scale observations can, in principle, be translated to the 
the constrained region of parameter space containing $G_{i}$ and $m_{\chi}$.
It is obvious from Fig.~\ref{si_clust} that for a given DM mass the variations of 
the constrained couplings $G_{i}$ are within $\mathcal{O}$(10) among different types of couplings
except for the cases of scalar-pseudoscalar (SP) and alternative tensor ($\tilde{\rm T}$) interactions where the 
calculated bounds are found to be somewhat relaxed to much higher values of couplings.
It can be inferred from Fig.~\ref{si_clust} that the obtained values of DM effective self couplings
constrained by the cluster observations for each considered DM mass scale are much higher 
by a few orders of magnitude than those of DM effective couplings with SM fermions~\ref{Beltran:2008xg,
Zheng:2010js} required by Planck relic density constraints.

\section{Summary and Conclusion \label{sec6}}

The concept of DM self interaction was first proposed to alleviate myriads of astrophysical problems. 
It also plays very significant role for the formation and growth of structures 
of the dwarf galaxies as well as of the galaxy clusters.
In this work, we discuss the possibility to constrain DM self interaction using the effective 
interactions among self interacting DM particles. For this we have chosen a 
general Dirac fermionic DM and considered 
all possible effective 4-fermion interaction operators among such DM particles. 
We have found nine possible types of such effective self interactions 
which yield independent forms for DM self scattering cross sections. 
These interactions include scalar (S), pseudoscalar (P), scalar-pseudoscalar (SP), 
vector (V), axialvector (A), vector-axialvector (VA), tensor (T), 
alternative tensor ($\tilde{\rm T}$) and chiral (C) types.
Using these effective interactions for the calculation of DM self interaction cross sections, 
it is found DM that for each type of those interactions, only three parameters, namely,
DM effective self coupling ($G_{i}$), DM mass ($m_{\chi}$) and relative velocity ($v_{rel}$) 
are required. A detailed study on how these parameters depend on each other has been performed.
From such study it has been shown that for a given value of DM mass and self interaction, 
the DM effective self couplings ($G_{i}, i= {\rm V, A, P, T, C}$) do not change considerably 
for a broad range of DM relative velocity
in the cases of vector, axialvector, pseudoscalar, tensor and chiral types of
interaction. On the other hand, for the cases of scalar, scalar-pseudoscalar, vector-axialvector and
alternative tensor interactions, the DM effective self couplings 
($G_{i}, i= {\rm S, SP, VA},\tilde{\rm T}$) decrease with the increment of $v_{\rm rel}$.
However, for any type of interaction there is a rapid decrement of $G_{i}$ with increasing $v_{\rm rel}$
near the relativistic regime of DM self interaction. This happens possibly be due to the strong
enhancement of self interaction cross section which, in turn, results a reduction of self coupling
near such very high relative velocity. 
Likewise, the self couplings, $G_{i}$ for a given
value of ${\sigma \over m_{\chi}}$ fall off with increasing DM mass ($m_{\chi}$). 
These decrements of $G_{i}$ with increasing $m_{\chi}$ follow a power law behaviour 
with index of power law being $-0.5$. 
We have also shown the allowed values of $G_{i}$ for a range of $m_{\chi}$ for two extreme cases 
of $v_{\rm rel}$, namely, for $v_{\rm rel} = 0$ (non-relativistic collision) and $0.999c$ (relativistic 
collision). From such study it can be inferred that the constraints on $G_{i}$ for different
$m_{\chi}$ become tighter when DM particles collide (or interact) with each other relativistically
while such constraints are relaxed for non-relativistic interaction between DM particles. 
However, the scale of such relaxations precisely depends on the nature of self couplings. 
The relaxation is comparatively low for tensor interaction among all interactions whereas
for the cases of scalar, scalar-pseudoscalar, vector-axialvector and alternative tensor
interactions it becomes so many orders of magnitude large that there is no preferred limit
for such non-relativistic interaction.
In addition, it is to be mentioned in this context that these constrained values of $G_{i}$
shift towards higher values if DM self interaction becomes stronger (i.e., self-scattering 
cross section is higher) keeping other parameters constant in this effective model and 
it also follows conversely.

In order to get parameter space consistent with various astrophysical bounds for self interaction,
we require the characteristic velocities for different size halos which may, in turn, determine the 
DM velocity distributions. 
Therefore, in order to derive the limits on $G_{i}$ for different $m_{\chi}$ from the observed results
of galaxy clusters, we need to perform velocity-averaging of the obtained
velocity dependent self scattering cross sections
over a Maxwellian velocity distribution with most probable speed being 
the characteristic cluster velocity, $v_0 \sim 10^3$ km/s. Finally few observed constraints 
on cluster scales are put 
into such resulting velocity-averaged cross section terms ($\langle \sigma \rangle$).
We have found that for the considered choice of $m_{\chi}$ being 10 MeV to 1 TeV, 
the values of $G_{i}, i= {\rm V, S, P, A, VA, T, C}$ roughly lie between $\sim 10^3$ GeV$^{-2}$ to 
$\sim 1$ GeV$^{-2}$ and those for $G_{i}, i= {\rm SP}, \tilde{\rm T}$ become
$\sim 10^5$ GeV$^{-2}$ to $\sim 10^3$ GeV$^{-2}$ consistent with ${\sigma \over m} = 1.5$ cm$^2$/g
from Abell 3827 cluster. On the other hand, for the requirement of 
${\sigma \over m} = 1.7\times10^{-4}$ cm$^2$/g from another analysis,
it requires $G_{i}, i= {\rm V, S, P, A, VA, T, C}$ to lie between $\sim 10$ GeV$^{-2}$ to 
$\sim 0.1$ GeV$^{-2}$ and $G_{i}, i= {\rm SP}, \tilde{\rm T}$ to take values 
between $\sim 10^3$ GeV$^{-2}$ and $\sim 10$ GeV$^{-2}$. Although the results are shown for a 
particular range of DM mass, the analysis can, in principle, be valid for any choice of DM mass.
Since effective coupling determines the fundamental energy scale, $\Lambda$ of the interaction
($G_{i} \sim {1\over \Lambda^2}$),
we can conclude from our study that the constrained value of such energy scale reduces
as the mass of DM fermion becomes higher.

In this work we thus have derived the self interaction parameters and their interrelations 
in an effective model for 4-point interaction among fermion WIMPs and 
further constrained the effective DM self couplings for cluster observational data.
The derived constraints on self couplings $G_{i}$ from self interaction are found out to be 
less severe by few orders of magnitude than those for the effective couplings among 
fermionic DM and SM fermions consistent with DM relic abundance bounds by Planck observation.
The general analysis performed in this study can also be extended for other DM self interaction limits
from other astrophysical sites such as from dwarf galaxies, Milky Way where the characteristic
DM velocity scale is smaller than that of clusters.
Alternatively since the DM particles are captured in the cores of massive astrophysical bodies like 
solar core, neutron stars etc. with capture rate prevailed by DM self interaction 
strength~\cite{Zentner:2009is}, this can also be used to constrain DM self couplings in this framework.

Although we have only studied the cases where each type of such interactions is 
the only effective interaction playing among the DM particles, we could have chosen the 
various combinations of all such interactions. This would constrain the 
the effective self couplings differently which can, in principle, be determined by combining the 
limits obtained in this work. Also there are scenarios with more than one WIMP on which effect of 
such self interaction constraints can be studied. Moreover the analysis performed for Dirac fermions
in this work can be extended for Majorana cases also. In addition, the yielded constraints in this 
study from self annihilation will restrict several particle physics models for fermionic DM such as
supersymmetric models with neutralino DM. 

%
%

\newpage

\section*{Acknowledgments}
\vspace{0.5cm}
KPM would like to thank Debasish Majumdar and Takashi Toma for their helpful suggestions.
KPM would also like to acknowledge Department of Atomic Energy (DAE, Govt. of India) for financial
assistance.

\end{document}